\documentclass[12pt]{article}
\usepackage{latexsym}
\newcommand{\beq}{\begin{equation}}
\newcommand{\eeq}[1]{\label{#1}\end{equation}}
\newcommand{\bea}{\begin{eqnarray}}
\newcommand{\eea}[1]{\label{#1}\end{eqnarray}}
\renewcommand{\Im}{{\rm Im}\,}

\begin{document}
\baselineskip 18pt
\begin{titlepage}
\hfill hep-th/0409210
\vspace{20pt}

\begin{center}
{\large\bf EFFECTIVE FIELD THEORY APPROACH TO COSMOLOGICAL INITIAL CONDITIONS: 
SELF-CONSISTENCY BOUNDS AND NON-GAUSSIANITIES}
\end{center}

\vspace{6pt}

\begin{center}
{\large M. Porrati} \vspace{20pt}

{\em Department of Physics\\ New York University\\ 4 Washington Place\\ 
New York NY 10003, USA}

\end{center}

\vspace{12pt}

\begin{center}
{\bf Abstract}
\end{center}
\begin{quotation}\noindent 
Effective Field Theory (EFT) is an efficient method for parametrizing unknown 
high energy physics effects on low energy data. When applied to 
time-dependent backgrounds, EFT must be supplemented with initial conditions.
In these proceedings, I briefly describe such approach, especially in the 
case of inflationary, almost-de Sitter backgrounds. I present certain 
self-consistency constraints that bound the size of possible 
deviations of the initial state from the standard thermal vacuum. I also 
estimate the maximum size of non-Gaussianities due to a non-thermal initial 
state which is compatible with all bounds.
These non-Gaussianities can be much larger than those due to nonlinearities in
the action describing single-scalar slow roll inflation. 
\end{quotation}
\vfill
 \hrule width 5.cm
\vskip 2.mm
{\small \noindent e-mail: massimo.porrati@nyu.edu}
\end{titlepage}
\section{EFT Approach to the Choice of Initial Conditions}

\subsection{Introduction}
The possibility of observing very high-energy, ``trans-Planckian'' 
physics in the cosmic 
microwave background radiation, thanks to the enormous stretch in proper
distance due to inflation, is one of the most exciting possibility for probing
string theory, or any other model of quantum gravity. As such, it has received
considerable attention, once the possibility was raised that these
effects could be as large as $H/M$, with $H$ the Hubble parameter during 
inflation, and $M$ the scale of new physics (e.g. the string scale).
A partial list of references is given in~\cite{p1}, on which this contribution
is largely based.

Due to our ignorance of the ultimate theory governing high-energy physics,
the most natural, model-independent approach to studying modifications to the
primordial power spectrum is effective field theory (EFT)~\cite{kkls,bch}. 
Using an EFT approach, the authors of~\cite{kkls} concluded that
the signature of any trans-Planckian modification of the standard inflationary
power spectrum is $O(H^2/M^2)$, well beyond the reach of observation even in 
the most favorable scenario ($H\sim 10^{14}\,\mbox{GeV}, M\sim 10^{16}\,
\mbox{Gev}$).

What was absent from e.g. 
ref.~\cite{kkls} was a {\em systematic} EFT approach to
initial conditions. That work presented convincing arguments against the 
(in)famous $\alpha$-vacua~\cite{alpha} of de Sitter space, but it did 
not give a complete parametrization of finite-energy non-thermal states.  

That parametrization was given in~\cite{sspds}, where the EFT approach was 
systematically extended to the choice of initial conditions. 
\subsection{The ``Initial'' State}
Let me review a suitably modified version of the approach of~\cite{sspds}.

The most important difference between~\cite{sspds} and other approaches is that
in~\cite{sspds} 
initial conditions for modes of all wavelengths are specified at the 
same initial time $t^*$. Other approaches give the initial conditions 
separately for each mode, at the 
time it crosses the horizon. The latter prescription is useful
in the context of inflationary cosmology, but it obscures the field-theoretical
meaning of the perturbation and/or initial condition: it does not easily
account the fact that after $t^*$ curvatures and energy densities are small, 
so the field theory is under control, and it does not easily
translate into an EFT language. The former prescription, instead, leads 
naturally to a simple classification of initial conditions in terms of 
{\em local} operators defined at the space-like boundary 
(i.e. initial surface) $t=t^*$.  

The prescription start by supplementing the EFT action describing all 
relevant low energy fields with a boundary term that encodes the standard 
thermal vacuum. To be concrete, we will work out the example of a massless 
scalar field in a time-dependent background. The 4d (bulk) action plus 
a 3d boundary term is
\bea
S&=&S_4+S_3, \qquad S_4=\int_{t\geq t^*}  
d^4 x \sqrt{-g} g^{\mu\nu}\partial_\mu \phi^* \partial_\nu \phi  \nonumber \\
S_3 &=& \int_{t=t^*} d^3 x \sqrt{\gamma(x)} 
\int_{t=t^*} d^3 y \sqrt{\gamma(y)} \phi^*(x) \kappa(x,y) \phi(y).
\eea{q1}
Here $\gamma_{ij}$ is the induced metric on the surface $t=t^*$. The role of
$S_3$ is to specify the wave functional for the scalar $\phi$ at $t=t^*$:
\beq
\Psi[\phi(x)]=\exp(iS_3[\phi]).
\eeq{q2}
Selecting an initial state for $\phi$ means in this language to choose a 
particular $\kappa(x,y)$. For instance, in de Sitter space with line element
\beq
ds^2= a(\eta)^2(-d\eta^2 + dx^idx^i), \;\;\; a(\eta)=-{1\over H\eta},
\;\;\; i=1,2,3, \;\;\; -\infty <
\eta <0,
\eeq{q2a} 
the standard thermal~\cite{bd} vacuum is obtained by choosing
\beq
\tilde{\kappa}(k)= -{k^2\eta^*\over 1-ik\eta^*}.
\eeq{q3}
Here, a tilde denotes the Fourier transform from space coordinates to 
co-moving momenta $k^i$ ($k\equiv\sqrt{k^ik^i}$) and $\eta^*$ is the initial 
(conformal) time.\footnote{From now on, $\eta$, $\eta^*$ will denote the
conformal time, $t$, $t^*$ will denote the synchronous proper time, and
$a()$ will always denote the scale factor.}
This expression for $\kappa$ makes clear that the choice of
such initial time is conventional, since a change in $\eta^*$ changes only 
$\kappa$, not the wave functional. From now on the standard vacuum functional 
will be called $|0\rangle$.
\subsection{Changing the Initial State}
Next, we want to find a convenient classification of changes in the initial 
state. This can be done by adding a new boundary term to the action:
$S\rightarrow S+\Delta S_3$. To determine $\Delta S_3$, we notice that, at any
finite time $t$ after $t^*$, we are insensitive to changes that only affect
very low
co-moving momenta $k$: co-moving momenta $k< H(t) a(t)$ correspond to
perturbations with super-horizon physical wavelength $\lambda_p>1/H(t)$, which
are unobservable at time $t$. So, since we are interested in changes that can 
be observed in the CMB of the present epoch, we have an IR cutoff naturally
built into the theory. This IR cutoff tells us that observable changes in 
the initial conditions can be parametrized by {\em local} operators:
\beq
\Delta S_3 = \sum_i \beta_i  M^{3-\Delta_i} 
\int_{t=t^*} d^3x \sqrt{\gamma} O^i.
\eeq{q4}
Here $O^i$ are operators of scaling 
dimension $\Delta_i$, $M$ is the high-energy 
cutoff of the EFT and the $\beta_i$'s are dimensionless parameters. 
The dimension $\Delta_i$ determines among other things how ``blue'' is the 
change in the power spectrum: the fractional change
in the power spectrum is proportional to $k^{\Delta_i-2}$. Since the
EFT makes sense only for $k<M$, operators of high conformal dimension do not 
significantly change the observable spectrum. So, the most significant 
observable changes in the primordial fluctuation spectrum are parametrized by 
a few local operators of low conformal dimension.

We just mentioned that the EFT needs a UV cutoff. This means that the operators
$O_i$ have to be suitably regulated at short distance. In other words, they 
are local only up to the cutoff scale $M$. As a simple example, consider the 
dimension-four operator $O^4=(\beta/M)(\partial_i\phi)^2$. It has to be smeared
at short distance, for instance by the replacement
\beq
\partial_i\phi\partial_i\phi \rightarrow 
\partial_i\phi f(-\partial^2/a^2(t^*)M^2)\partial_i\phi.
\eeq{q5}
Here $f(x)$ is a smooth function obeying  $f(x)=1$, for $x\leq 1-\epsilon$;
$f(x)=0$, for $x\geq 1+\epsilon$; $\epsilon$ is a small positive number.
The scale factor 
$a(t^*)$ appears because we want to cutoff at $M$ the {\em physical} momentum 
$k/a(t^*)$, not the co-moving momentum $k$.
\subsection{Power Spectrum}
As a first application, let us derive the change in the power 
spectrum of a minimally-coupled scalar field in de Sitter space,
induced by the operator $O^4$~\cite{sspds}. The change in initial 
conditions $\Delta S_3$, is equivalent to perturbing the Hamiltonian of the 
system by an instantaneous interaction $H_I=-\delta(t-t^*)\Delta S_3$. 
So, the perturbed power spectrum is
\beq
P(k)=\lim_{\eta\rightarrow 0^-} \langle |\phi(k,\eta)|^2\rangle =
\lim_{\eta\rightarrow 0^-}
\langle 0| \exp (-i\Delta S_3) |\phi(k,\eta)|^2  \exp (i\Delta S_3)|0\rangle.
\eeq{q6}
To first order in $\beta$, the change is
\beq
\delta P(k) = -i {\beta\over M}\int d^3x \langle 0| [O^4(x), 
|\phi(k,0)|^2 ]|0\rangle.
\eeq{q7}
This quantity is easily computed in terms of commutators of free fields in de
Sitter space, resulting in~\cite{sspds,p1}
\beq
\delta P(k) = -{\beta \over M} {H^2\over k^3} \Im [\phi^+(\eta^*,k)]^2 k^2
f(k^2/a^2(\eta^*)M^2).
\eeq{q8}
$H^2/ k^3$ is the unperturbed power spectrum, while
the canonically normalized, positive frequency solution of the 
free-field equations of motion is 
\beq
\phi^+(\eta,k)= {H\over \sqrt{2k^3}}(1-ik\eta)\exp(ik\eta).
\eeq{q9}
For $k\eta^* \sim 1$, the effect of $O^4$ on the power spectrum can be as
large as 
$\delta P/P \sim \beta H/M$, i.e. in the observable range when $\beta$ is 
$O(1)$. 
\section{Back-Reaction and Calculability Bounds}
Reference~\cite{sspds} does not take into account all effects due to the
back-reaction of the modified stress-energy tensor on the metric. 
Specifically, any change in the boundary conditions of the effective field 
theory generates 
modifications to the expectation value of the stress-energy tensor. These
modifications can become large near the (space-like) boundary hypersurface.
By requiring that the back-reaction remains under control, we shall get new 
bounds on the size of the parameters $\beta_i$.

From now on, unless otherwise stated, we will set $a(t^*)=1$ for ease of 
notation. 

Since gravity couples universally to matter through the stress-energy tensor,
any change in the expectation value of $T_\mu^\nu$ will back-react on the 
metric and change the background, that will be no longer a pure de Sitter 
space. In our formalism, the change in $\langle T_\mu^\nu \rangle $ 
to first order in the $\beta_i$'s is easily written as
\beq
\delta \langle T_\mu^\nu(t,x) \rangle =
-i \langle 0| [\Delta S_3, T_\mu^\nu (t,x)]|0\rangle.
\eeq{q10}

Notice that we are looking for a first-order change in 
$\langle T_\mu^\nu \rangle $. This is different from {\em second-order} effects
due to the change in the vacuum, considered elsewhere in the 
literature~\cite{st}, such as particle production \& c.
The change we are considering here is vanishingly small for times
$\Delta t=t-t^* \gg  1/M$, but it can be large for times $\Delta t$ of order 
$1/M$. A direct computation~\cite{p1} shows that under the 
perturbation $O^4$, $\langle T_0^0 \rangle $ does not change to first order 
in $\beta$, while  $\delta\langle T_i^i \rangle $ can be as large as
\beq
\delta \langle T_i^i(t,x) \rangle \approx \beta M^4g(\Delta t M).
\eeq{q11}
The exact form of the function $g(x)$ depends on the shape of the 
cutoff function $f$, 
but it is always $O(1)$ inside the region $x \sim 1$ and it vanishes
for $x\gg 1$.

Now, a change in the  pressure, 
$\delta p = \delta \langle T_i^i(t,x) \rangle $, implies a change in the 
Hubble constant: 
\beq
\delta \dot{H}= 4\pi G \delta p \approx  4\pi G \beta M^4, \qquad 
H={\dot{a}\over a}\; .
\eeq{q12}
Combined with the standard slow-roll conditions $\dot{H}=\epsilon H^2$, 
$\ddot{H}=2\epsilon\eta' H^3$,\footnote{Typically, 
$\epsilon,\eta' \leq 10^{-2}$.}
this equation, and the obvious estimate $\dot{g}(\Delta tM)|_{t\approx t^*} 
\sim M$, implies severe constraints on $\beta$:
\beq 
\beta \leq \epsilon {1\over 4\pi G M^2} {H^2\over M^2}, \qquad
\beta \leq 2\epsilon \eta' {1\over 4\pi G M^2} {H^3\over M^3}.
\eeq{q13} 
They can easily rule out the observability of any change in the power spectrum.
\subsection{What The Bounds Mean}
Since the back-reaction effect we found is limited to a short time of order 
$1/M$ after $t^*$, one may think that it should be possible to relax 
the slow roll conditions for such a short time without any observable 
consequence on the power spectrum, except for those modes that cross the 
horizon within a time $\Delta t$ after $t^*$. This is not so, because the 
correct way of thinking about $\Delta S_3$ is as a parametrization of all 
changes that happened at {\em any} time before $t^*$. In other words, all
modes that cross the horizon {\em before} $t^*$ can (and are) affected 
significantly. Consider in particular the case of single-scalar slow roll 
inflation. The scalar fluctuations of the metric are 
described by a gauge invariant variable $v$, whose action is that of a 
minimally coupled free scalar with a time-dependent mass term
(see \cite{mfb} and references therein) 
\bea
S&=& \int d^3x dt a^3\left[-\dot{v}^2 + 
a^{-2}(\partial_i v)^2 + {\cal M}^2 v^2\right],\nonumber \\ 
{\cal M}^2&=&-3\dot{H} + {3H \ddot{H}
\over 2 \dot{H}} -{2\ddot{H}\over H} + {2\dot{H}^2\over H^2} -
{\ddot{H}^2\over 4\dot{H}^2} + 
{\stackrel{\, ...}{H}\over 2\dot{H}} \; .
\eea{q14}  
The change in $\dot{H}$ is confined to 
within a time $\Delta t \sim 1/M$, so, in 
looking at modes of wavelength longer than the cutoff $1/M$, 
we can approximate its effect by replacing the time-dependent terms induced in
${\cal M}$ by the back-reaction with $\delta(t-t^*) \delta{\cal M}/M$. 
This is the same as adding a new operator in $\Delta S_3$: 
$O=\int d^3x (\delta {\cal M}/M)v^2$. When $\dot{H} \gg \epsilon H^2$ the
new induced boundary term ${\delta {\cal M}}/M$ is $O(M)$, and
it changes the spectrum as~\cite{p2}
\beq
{\delta P(k)\over P(k)}= {M\over H},\qquad\mbox{for } k\eta^* \ll 1.
\eeq{q15}
This change is {\em never} a small perturbation of the de Sitter space result, 
so we must satisfy the slow-roll condition $\dot{H}\ll\epsilon H^2$ 
and re-evaluate 
the change in  ${\delta {\cal M}}/M$. It can be estimated as~\cite{p2} 
$\sim \beta M^5G/\epsilon H^2$. By asking again that the change in
the power spectrum $\delta P(k)/P(k)$ is not greater than $O(1)$ we find 
 estimate $\beta \leq O(\epsilon H^3/GM^5)\sim 10^{-2}$.
If we ask that the change is smaller than the one computed at tree level 
[Eq.~(\ref{q8})] we must have ${\stackrel{\, ...}{H}}\ll \epsilon\eta\eta'H^4$,
where $\eta'$ is another slow-roll parameter of magnitude comparable to
$\epsilon$ and $\eta$. This gives a very strong estimate:  
$\beta \leq O(\epsilon\eta\eta' H^4/GM^6)$.
 
This method for arriving at a bound is more involved than that leading to 
Eq.~(\ref{q13}), but it is more satisfying. The slow roll expansion is not
assumed to be valid at times infinitesimally close to $t^*$, and the meaning of
the bound is clearer: if we do not impose it, then the change in the power 
spectrum computed in ref.~\cite{sspds} and by Eq.~(\ref{q8}) is in reality
sub-dominant compared to that due to the back-reaction on the metric.
\section{Non-Gaussianities}
Other changes to the initial state exist, that give a potentially 
observable signal while being compatible with back-reaction bounds.
One such change is a non-Gaussianity {\em in the initial conditions}.
It is induced by the boundary action\footnote{In this section we revert to
using conformal time and $a(\eta^*)\neq 1$.}
\beq
\Delta S_3= \int_{\eta=\eta^*} d^3x a^3 \lambda v^3.
\eeq{q16}
To first order in $\lambda$, this term induces a three-point function 
for $v$~\cite{p2}:
\beq
\langle \tilde{v}(k_1)  \tilde{v}(k_2) \tilde{v}(k_3) \rangle
= -i \lambda \int d^3x a^3 
\langle 0 | [v^3(\eta^*,x),\tilde{v}(0,k_1)  \tilde{v}(0,k_2) 
\tilde{v}(0,k_3)]|0\rangle. 
\eeq{q17}
A short calculation gives
\bea
\langle \tilde{v}(k_1)  \tilde{v}(k_2) \tilde{v}(k_3) \rangle &=&
-(2\pi)^3\delta^3(k_1+k_2+k_3) {\lambda H^3\over 4} \sum_{i>j} 
k_i^{-3}k_j^{-3}, \;\;\; |k_i|\eta^* \ll 1, \nonumber \\
&\approx & 0 , \qquad  |k_i|\eta^* \gg 1.
\eea{q18}
This functional dependence is similar to the universal non-Gaussianities due 
to the bulk gravitational action~\cite{m}; except for the cutoff effect at 
$|k_i|\eta^* \sim 1$, which is absent in the bulk effect. 
This cutoff is another illustration of the fact that
the boundary term $\Delta S_3$ is physically equivalent to changing the 
evolution of space-time at all times before $\eta^*$: wavelength that are still
inside the horizon at $\eta^*$ are not affected significantly by past history,
due to the exponential expansion of the background.

Back-reaction effects are of two types: one is the second-order change in
$\delta\langle T_\mu^\nu\rangle = -\langle 0 | [\Delta S_3, [ \Delta S_3, 
T_\mu^\nu]]|0\rangle $. The other arises from the first-order interference
between $\Delta S_3$ and the cubic terms in $v$ present in 
$T_\mu^\nu$. The worst-case scenario estimate for these terms is~\cite{p2}
\beq
\delta\langle T_\mu^\nu\rangle \sim \lambda^2 M^4 + 
\sqrt{\epsilon} |\lambda| {M^5\over M_{Pl}} \ll 
O\left(\epsilon \eta' {M_{Pl}^2 H^4\over M^2}\right).
\eeq{q19}
These bounds allow for a $\lambda$ as large as
$O(\sqrt{\epsilon\eta} M_{Pl}H^2/M^3)$. 
This translates into a coefficient for the  non-Gaussianity 
Eq.~(\ref{q18}) as large as $O(\sqrt{\epsilon\eta} M_{Pl}H^5/M^3)$. 
To compare with Maldacena's 
result~\cite{m}, we convert his variable $\zeta$ into $v$ and we arrive to a
non-Gaussianity coefficient $\sqrt{\epsilon}H^4/M_{Pl}$. The ratio of
our coefficient to Maldacena's is
\beq
{\sqrt{\epsilon\eta} M_{Pl}H^5/M^3\over \sqrt{\epsilon}H^4/M_{Pl}}=
\sqrt{\eta}{M^2_{Pl}H\over M^3}.
\eeq{q20}
This number can be very large, easily larger than $10^2$. Thus,
the non-Gaussianity due initial conditions can be observably large, provided, 
of course, that the initial time $\eta^*$
is no more than about 60 e-foldings away from
the end of inflation. Otherwise, the cutoff at $|k| \sim 1/\eta^*$ would wash
out all effects on any observable, sub-horizon fluctuation.
\subsection*{Acknowledgments}
I would like to thank E. Mottola and G. Shiu for interesting discussions, and 
acknowledge the Aspen Center for Physics, where part of this work was 
completed, for its hospitality. Work supported in part by NSF grant 
PHY-0245068.


\begin{thebibliography}{0}
\bibitem{p1} M.~Porrati,
Phys.\ Lett.\ B {\bf 596}, 306 (2004)
[arXiv:hep-th/0402038].
\bibitem{kkls}
N.~Kaloper, M.~Kleban, A.~E.~Lawrence and S.~Shenker,
Phys.\ Rev.\ D {\bf 66}, 123510 (2002)
[arXiv:hep-th/0201158];
N.~Kaloper, M.~Kleban, A.~Lawrence, S.~Shenker and L.~Susskind,
JHEP {\bf 0211}, 037 (2002)
[arXiv:hep-th/0209231].
\bibitem{bch} 
C.~P.~Burgess, J.~M.~Cline, F.~Lemieux and R.~Holman,
JHEP {\bf 0302}, 048 (2003)
[arXiv:hep-th/0210233];
C.~P.~Burgess, J.~M.~Cline and R.~Holman,
JCAP {\bf 0310}, 004 (2003)
[arXiv:hep-th/0306079];
C.~P.~Burgess, J.~Cline, F.~Lemieux and R.~Holman,
arXiv:astro-ph/0306236.
\bibitem{alpha}
N.~A.~Chernikov and E.~A.~Tagirov,
Annales Poincar\'e Phys.\ Theor.\ A {\bf 9} 109 (1968);
E.~A.~Tagirov,
Annals Phys.\  {\bf 76}, 561 (1973); 
E.~Mottola,
Phys.\ Rev.\ D {\bf 31}, 754 (1985);
B.~Allen,
Phys.\ Rev.\ D {\bf 32}, 3136 (1985);
J.~G\'eh\'eniau and C.~Schomblond, Bull. Class. Sci. Acad. Roy. Belg. 
{\bf 54}, 1147 (1968); C.~Schomblond and P.~Spindel, Annales Poincar\'e Phys.\
Theor. \ A {\bf 25}, 67 (1976).  
\bibitem{sspds}
K.~Schalm, G.~Shiu and J.~P.~van der Schaar,
arXiv:hep-th/0401164.
\bibitem{bd} N.~D.~Birrell and P.~C.~W.~Davies,
{\it Quantum Fields In Curved Space}, Cambridge Univ. Press (1982).
\bibitem{st}  A.~A.~Starobinsky and I.~I.~Tkachev,
JETP Lett. {\bf 76}, 235 (2002)
[Pisma Zh.\ Eksp.\ Teor.\ Fiz.\  {\bf 76}, 291 (2002)]
[arXiv:astro-ph/0207572];
T.~Tanaka,
arXiv:astro-ph/0012431;
M.~Giovannini,
Class.\ Quant.\ Grav.\  {\bf 20}, 5455 (2003)
[arXiv:hep-th/0308066].
\bibitem{mfb} V.~F.~Mukhanov, H.~A.~Feldman and R.~H.~Brandenberger,
Phys.\ Rept.\  {\bf 215}, 203 (1992).
\bibitem{p2} M. Porrati, in preparation.
\bibitem{m}
J.~Maldacena,
JHEP {\bf 0305}, 013 (2003)
[arXiv:astro-ph/0210603].
\end{thebibliography}
\end{document}